\documentclass[twocolumn, a4paper, superscriptaddress, pra, showpacs]{revtex4-1}
\usepackage{graphicx, epsfig, fancybox}
\usepackage{amssymb,amsmath}
\usepackage{color}

\usepackage{times}

\def \del{\partial}    
\def \hf{\tfrac{1}{2}}

\newcommand{\ket}[1]{\left|{#1}\right.\rangle}

\begin{document}

\title{Spin modulation instabilities and phase separation dynamics in trapped two-component Bose condensates}

\author{Ivana Vidanovi\'c}

 \affiliation{Scientific Computing Laboratory, Institute of Physics Belgrade,
 University of Belgrade, Pregrevica 118, 11080 Belgrade, Serbia}

 \affiliation{Institut f\"ur Theoretische Physik, Goethe-Universit\"at, 60438 Frankfurt/Main, Germany}

\affiliation{Max Planck Institute for the Physics of Complex Systems, N\"othnitzer Str.~38, 01187 Dresden, Germany}



 \author{N.~J.~van~Druten}

\affiliation{Van der Waals-Zeeman Institute, University of Amsterdam, Science
   Park 904, 1098 XH Amsterdam, The Netherlands}

 \author{Masudul Haque}

\affiliation{Max Planck Institute for the Physics of Complex Systems, N\"othnitzer Str.~38, 01187 Dresden, Germany}

\begin{abstract}

In the study of trapped two-component Bose gases, a widely used dynamical
protocol is to start from the ground state of a one-component condensate and
then switch half the atoms into another hyperfine state.  The slightly
different intra-component and inter-component interactions can then lead to
highly nontrivial dynamics.  We study and classify the possible subsequent
dynamics, over a wide variety of parameters spanned by the trap strength and
by the inter- to intra-component interaction ratio.  A stability analysis
suited to the trapped situation provides us with a framework to explain the
various types of dynamics in different regimes.

\end{abstract}




\maketitle

\section{Introduction} 

Two-component Bose-Einstein condensates (BECs) are increasingly appreciated as
a rich and versatile source of intricate non-equilibrium pattern dynamics
phenomena.  In addition to experimental
observations \cite{MatthewsWiemanCornell_PRL98, HallWiemanCornell_PRL98,
MiesnerStamper-KurnStenger_PRL99, ModugnoRiboliRoatiInguscio_PRL02,
LewadowskiCornell_PRL02, MertesKevrekidisHall_PRL07, PappWieman_PRL08,
vanDruten_expt, HamnerEngelsHoefer_PRL11, Treutlein_NatPhys09,
AndersonTicknorSidorovHall_PRA09, EgorovDrummondSidorov_PRA11, MyattCornellWieman_PRL97}, pattern
dynamics in two-component BECs has attracted significant theoretical interest
(see, e.g., \cite{HoShenoy_PRL96, LawPuBigelowEberly_PRL97,
BuschCiracGarciaZoller_PRA97, PuBigelow_PRL98, Timmermans_PRL98,
GrahamWalls_PRA98, SinatraFedichevCastinDalibardShlyapnikov_PRL99,
KasamatsuTsubota_PRL04, KasamatsuTsubota_PRA06, Ronen_PRA08,
NavarroKevrekidis_PRA09, WenLiuCaiZhangHu_PRA12, ZurekDavis_PRL11,
LiTreutleinSinatra_EPJB09, BalazNicolin_PRA12}
and citations in \cite{NavarroKevrekidis_PRA09}).

In a number of two-component BEC experiments reported over more than a decade,
a standard technique has been to start from the equilibrium state of a
single-component BEC, e.g., populating a single hyperfine state of $^{87}$Rb,
and then using a $\pi/2$ pulse to switch half the atoms to a different
hyperfine state \cite{HallWiemanCornell_PRL98, MatthewsWiemanCornell_PRL98, 
MiesnerStamper-KurnStenger_PRL99, LewadowskiCornell_PRL02,
MertesKevrekidisHall_PRL07, vanDruten_expt, EgorovDrummondSidorov_PRA11,
AndersonTicknorSidorovHall_PRA09, Treutlein_NatPhys09}.  This results in a
binary condensate where the two intra-species interactions ($g_{11}$ and
$g_{22}$) and one inter-species interaction ($g_{12}$) are all slightly
different from each other, but the starting state is the ground state
determined by $g_{11}$ alone.  Since it has been realized several times in
several different laboratory setups, this is a paradigm non-equilibrium
initial state for binary condensate dynamics.  A thorough and general analysis
of the dynamics subsequent to such a $\pi/2$ pulse is thus clearly important.
In this article we present such an analysis, clarifying the combined role of
the inter-species interaction ($g_{12}$) and the strength $\lambda$ of the
trapping potential.  We provide a stability analysis mapping out regions of
the $\lambda$--$g_{12}$ parameter space hosting different types of dynamics.
Since it is now routine to monitor real-time dynamics in such experiments
(e.g.~\cite{vanDruten_expt}), we also directly analyze the real-time evolution
after a $\pi/2$ pulse.

It is widely known that the ground state of a uniform two-species BEC is phase
separated or miscible depending on whether or not the inter-species repulsion
dominates over the self-repulsions of the two species, i.e., if
\begin{equation} 
g_{11} g_{22} ~<~ \left(g_{12}\right)^2
\label{eq:cond}
\end{equation}
then the ground state is phase separated \cite{HoShenoy_PRL96}.  This
criterion is also a key ingredient in understanding dynamical features such as
pattern dynamics in the density difference between the two species --- such
``spin patterns'' emerge when the phase separation condition is satisfied.
This can be understood as the onset of a modulation instability
\cite{Timmermans_PRL98, KasamatsuTsubota_PRL04, KasamatsuTsubota_PRA06},
identified by the appearance of an unstable mode in the excitation spectrum
around a reference stationary state.  For a \emph{homogeneous} situation,
linear stability analysis shows that modulation instability sets in when the
condition of Eq.~(\ref{eq:cond}) is satisfied \cite{Timmermans_PRL98,
  KasamatsuTsubota_PRL04, KasamatsuTsubota_PRA06} .

The situation is different in the presence of a trapping potential.  Phase
separation in the ground state, as well as the appearance of modulation
instability when starting from a mixed state, now requires larger
inter-species repulsion \cite{NavarroKevrekidis_PRA09,
WenLiuCaiZhangHu_PRA12}.  This suggests that the region of parameter space
where pattern dynamics occurs also depends on the trap.  A trap is almost
always present in cold-atom experiments, and it is easy to imagine experiments
where the trapping potential is not extremely shallow but varies between tight
and shallow limits.  It is thus necessary to examine the relevance of
Eq.~(\ref{eq:cond}) for trapped binary BECs.  To this end, we explore
different trap strengths spanning several orders of magnitude, and identify
the appropriate extensions of Eq.~(\ref{eq:cond}) for the type of spin
dynamics resulting from the $\pi/2$ protocol described above.

We focus on the effects of two parameters.  First, we study effects of
changing cross-species interaction $g_{12}$, thus generalizing
Eq.~\eqref{eq:cond} for trapped situations.  Second, we explore the role of
the relative strength of the trap with respect to the interactions.  Our
analysis, performed for a one-dimensional (1D) geometry, sheds light on the
situation where $g_{11}$ and $g_{22}$ are close but unequal: (a) the stability
analysis is performed for $g_{11}=g_{22}$ and their difference serves only to
select appropriate instability modes; (b) the simulations are performed with
$g_{22}/g_{11}=1.01$.

In Section \ref{sec_model}, we introduce the formalism and geometry.  In
Section \ref{sec_stability_analysis}, we show results from a linear stability
analysis for a sequence of trap strengths, and identify and analyze relevant
modulation instabilities.  Through an analysis of unstable modes, we present a
classification of the parameter space into dynamically distinct regions, in
relation to the prototypical initial state explained above.  This may be
regarded as a dynamical ``phase diagram''.  A remarkable aspect is that the
``phase transition'' line most relevant to spin pattern dynamics does not
arise from the first modulation instability (studied in
Ref.~\cite{NavarroKevrekidis_PRA09}).  This first instability mode is
antisymmetric in space, and as a result is not naturally excited in a
symmetric trap with symmetric initial conditions.  Complex dynamics (not due
to collective modes but rather due to modulation instability) is generated
only when the first \emph{spatially symmetric} mode becomes unstable, which
occurs at a higher value of $g_{12}$.

In Section \ref{sec_timeEvolution} we provide a relatively detailed account of
the time evolution.  For each trap strength $\lambda$, for values of $g_{12}$
not much larger than $g_{11}$, we observe simple collective modes.  Above a
threshold value of $g_{12}$, the oscillation amplitude becomes sharply
stronger, and at the same time the motion becomes notably aperiodic, signaling
that the dynamics is more complex than a combination of a few modes.
Dynamical spin patterns start appearing at this stage and become more
pronounced as $g_{12}$ is increased further.  The threshold value at which the
dynamics changes sharply corresponds to the second modulation instability line
rather than the first, as we demonstrate through careful choice of parameters
in each region of the phase diagram derived from stability analysis.

Some further connections between the stability analysis and dynamical
features, relating to the length scale of generated patterns, appear
in Section \ref{sec_lengthscales}.  In the concluding
Section \ref{sec_conclusions} we place our results in context and point out
open questions.

\section{Geometry and  formalism} 
\label{sec_model}

The relevant time-resolved experiments have been performed in both quasi-1D
geometries (highly elongated traps with strong radial
trapping) \cite{vanDruten_expt} and in a 3D BEC of cylindrical symmetry with
the radial variable playing analogous role as the 1D
coordinate \cite{HallWiemanCornell_PRL98, MertesKevrekidisHall_PRL07}.  Since
the basic phenomena are very similar, we expect the same theoretical framework
to describe the essential features of each case.  For definiteness, in this
work we show results for 1D geometry.  We expect the general picture emerging
from this work to be qualitatively true also for other geometries exhibiting
the same type of spin dynamics.

We describe the dynamics in the mean field framework at zero temperature,
i.e., by two coupled Gross-Pitaevskii equations \cite{pitaevskii-jetp13,
gross-nc20, SalasnichParolaReatto_PRA02}:
\begin{eqnarray}
\hspace{-2mm}i\del_t \psi_1 &=& \bigg( -\hf\del^2_x +\hf\lambda^2x^2
+g_{11}|\psi_1|^2   +g_{12}|\psi_2|^2   \bigg)  \psi_1\,,\label{eq:gp1}
\\
\hspace{-2mm}i\del_t \psi_2 &=& \bigg( -\hf\del^2_x +\hf\lambda^2x^2
+g_{12}|\psi_1|^2  +g_{22}|\psi_2|^2   \bigg)  \psi_2\,.\label{eq:gp2}
\end{eqnarray}
Condensate wave functions $\psi_1(x,t)$ and $\psi_2(x,t)$ are normalized to
unity, and $\lambda$ is the strength of the harmonic trap.  Factors of
particle number and radial trapping frequency are absorbed as appropriate into
the effective 1D interaction parameters $g_{ij}$ \cite{Olshanii_PRL98,
SalasnichParolaReatto_PRA02, vanDruten_expt}.
We consider purely non-dissipative dynamics, i.e., we do not attempt to model
experimental loss rates with a phenomenological dissipative term as done in,
e.g., Refs.\ \cite{MertesKevrekidisHall_PRL07, vanDruten_expt,
AndersonTicknorSidorovHall_PRA09}.

The equations above are in dimensionless form because we measure lengths in
units of trap oscillator length and time in units of inverse trapping
frequency, for a hypothetical trap of unit strength ($\lambda=1$).  The scale
for trap strengths is itself fixed by imposing $g_{11}=1$.  With this
convention, small values of $\lambda$ correspond to a BEC in the Thomas-Fermi
limit.  For comparison, we note that the parameters of the experiment of
Ref.~\cite{vanDruten_expt} corresponds to $\lambda$ of order
$10^{-5}$ in these units.
%
%
Of course one can switch between different units via the
transformation: $ x \rightarrow x/l$, $ t\rightarrow t/l^2$, $\lambda
\rightarrow \lambda l^2$, $g \rightarrow g l$, and $\psi \rightarrow
\psi \sqrt{l}$, where $l$ is an appropriately chosen scale.

The initial state after a  $\pi/2$ pulse involves both components occupying the
ground state of a single-component system of interaction $2g_{11}$, because the
atoms were all in the first hyperfine state before the pulse.  We model this
initial situation as a two-component BEC with $g_{11}=g_{22}=g_{12}$.  The
$\pi/2$ pulse may then be regarded as a sudden change (a \emph{quantum
quench} \cite{PolkovnikovSengupta_RMP11}) of the interaction parameters
$g_{22}$ and $g_{12}$.

We use $g_{11}= g_{22}$ for the stability analysis of
Section \ref{sec_stability_analysis}.  For the explicit time evolution
reported in Section \ref{sec_timeEvolution}, we use $g_{11}$ and $g_{22}$
values close but unequal: $g_{11}=1$, $g_{22}=1.01$.
This choice of close values is convenient for illustrating the structure of
the phase diagram, especially for shallow traps.  In rubidium experiments the
difference between $g_{11}$ and $g_{22}$ is somewhat larger (in the common
case using $^{87}$Rb hyperfine states $\ket{1}=\ket{F=1,m_F=-1}$ and
$\ket{2}=\ket{F=2,m_F=1}$); however our insights should be relevant to a broad
regime of possible experiments.  A full exploration of the regime of arbitrary
differences ($g_{11}-g_{22}$) remains an open task, beyond the scope of the
present manuscript.

Numerical simulations presented in Section \ref{sec_timeEvolution} were
performed using a semi-implicit Crank-Nicolson method \cite{sicn,
IvanaAntun_2012}.

\begin{figure}[!tb]
\centering  \includegraphics[width= 0.99\columnwidth]{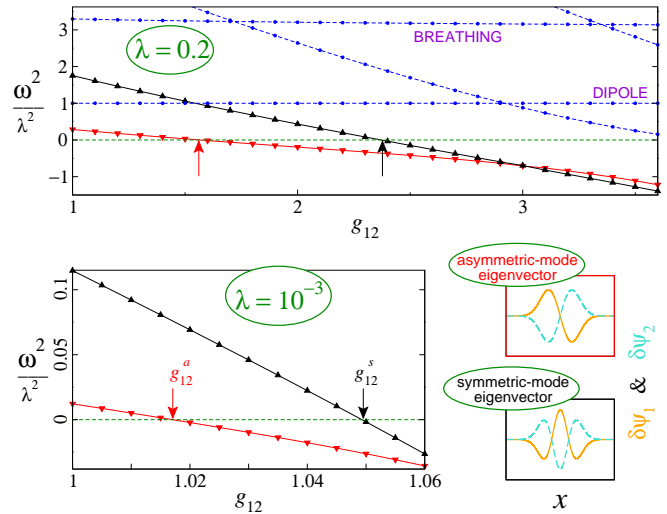}
\caption{ \label{fig:Fig3}  (Color online.) 
Results from stability analysis.  Squared eigenvalues $\omega^2$ of the
stability matrix $\mathcal{M}$ are plotted against $g_{12}$, for a tight trap
(top) and for a shallow trap (bottom left). The arrows show the values of
$g_{12}$ for onset of the two instabilities, namely $g_{12}^a$ (onset of
spatially antisymmetric modulation instability) and $g_{12}^s$ (onset of
spatially symmetric instability).  Typical eigenvectors corresponding to these
two modes are shown in the panels on lower right.
}
\end{figure}

\section{Stability analysis and  dynamical ``phase diagram''}  \label{sec_stability_analysis}

We provide in this section a stability analysis for $g_{11}=g_{22}$ that maps
out the regions of $\lambda$-$g_{12}$ parameter space which support pattern
formation instabilities.

Ideally, one might like to perform a stability analysis around the initial
state.  However, in contrast to the homogeneous case \cite{Timmermans_PRL98},
we are faced with the situation that the initial state is not a stationary
state of the final Hamiltonian.  The choice of reference state is therefore a
somewhat subtle aspect of the present analysis.

We use as reference state $\psi_{0}(x)$ the lowest-energy spatially symmetric
stationary state of the case $g_{11} = g_{22}$, with parameter $g_{12}$ set to
its final value.  (For large $g_{12}$, this is not the ground state for these
parameters, which is phase-separated.)  This reference state has the advantage
of looking relatively similar to our actual initial state (two components
totally overlapping in space), and of being a stationary state of the
Hamiltonian for which we analyze linear stability.  Our reference state can be
regarded as placing both components in the single-component ground state for
interaction $g_{11}+g_{12}$.  We are not aware of a suitable stationary state
even more similar to the actual initial state.  We will see that our stability
analysis around this reference state will predict remarkably well the main
observed time-evolution features described in Section \ref{sec_timeEvolution}.

Note that it is not natural to use $g_{11} \neq g_{22}$, because stationary
states for such a case typically do not overlap completely in space.  Instead,
in our approach the difference between $g_{11}$ and $g_{22}$ will play the
important role of selecting certain instability modes over others.  For this
reason, inferences from the present analysis apply only to small relative
differences between $g_{11}$ and $g_{22}$.

We linearize Eqs.~(\ref{eq:gp1}) and (\ref{eq:gp2}) around the  reference
stationary state $\psi_{0}(x)$: 
\begin{eqnarray}
\psi_1(x, t) = \left[ \psi_{0}(x) + \delta\psi_1(x,t) \right] \exp(-i \mu t),\nonumber\\
\psi_2(x, t) = \left[ \psi_{0}(x) + \delta\psi_2(x,t) \right] \exp(-i \mu t),
\end{eqnarray}
where $\mu$ is the chemical potential corresponding to the reference state.
By keeping only terms of the first order in $\delta\psi_1(x,t)$ and
$\delta\psi_2(x,t)$, we obtain a system of linear equations which can be cast
in the form:
\begin{equation}
\del_t^2
\begin{pmatrix} \delta\psi_1+  \delta\psi_1^* \\ \delta\psi_2 + \delta\psi_2^*\end{pmatrix}
+\mathcal{M} 
\begin{pmatrix} \delta\psi_1 +  \delta\psi_1^* \\ \delta\psi_2 + \delta\psi_2^*\end{pmatrix}~=~ 0.
\end{equation}
Here $\mathcal{M}$ is a matrix differential operator which, upon
discretization or upon expansion in a set of orthogonal functions, becomes the
so-called stability matrix (e.g., \cite{PuBigelow_PRL98,
LawPuBigelowEberly_PRL97}).  We analyze below the eigenmodes of the stability
matrix, which we have obtained by numerically calculating the reference
stationary state $\psi_{0}(x)$ and expanding in the basis of harmonic trap
(non-interacting) eigenstates.

Since we use $g_{11} = g_{22}$ for the stability analysis, eigenmodes will
have well-defined ``species parity'', i.e. will all be either even
[$\delta\psi_1(x,t) = \delta\psi_2(x,t)$] or odd [$\delta\psi_1(x,t) =
-\delta\psi_2(x,t)$] with respect to the interchange of species.
Even modes describe in-phase motion of the two components and simply
correspond to the excitation spectrum of a single-component BEC with
interaction constant $g_{11}+g_{12}$.  Odd modes are more interesting --- they
describe out-of-phase motion of two components and are therefore reflected in
the spin dynamics.  Additionally, due to the spatial inversion symmetry
$x\rightarrow -x$, the solutions will also have well-defined spatial parity,
and we can distinguish spatially symmetric and antisymmetric modes.

Typical eigenspectra are presented in Fig.~{\ref{fig:Fig3}}.  In the case of a
tight trap $\lambda = 0.2$, we notice two modes whose frequencies are nearly
constant.  These are even modes encoding single-component or in-phase physics.
The lower one is the dipole (Kohn) mode with frequency equal to the trap
frequency $\lambda$.  The second nearly-constant mode is the breathing mode,
which for elongated traps takes value close to $\omega^2 = 3 \lambda^2$.  The
breathing mode (oscillations of cloud size) is visible in the plots of
Fig.~\ref{fig:Fig1} (Section \ref{sec_timeEvolution}) as a fast oscillation of
the total condensate widths.

\begin{figure}[!tb]
\centering
\includegraphics[width=\columnwidth]{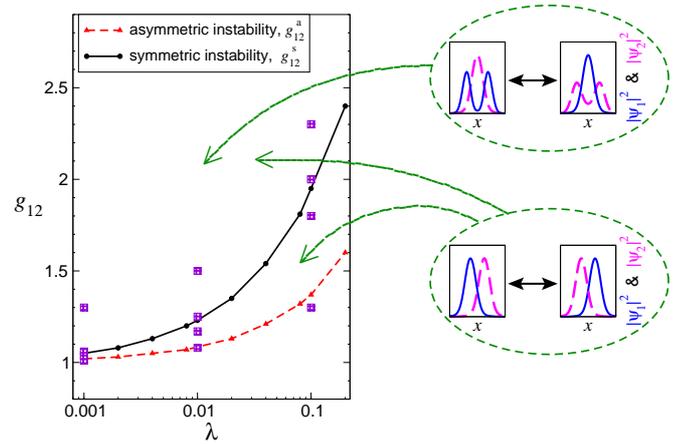}
\caption{ \label{fig:Fig4}  (Color online.) 
The dynamical ``phase diagram'' showing the critical values of $g_{12}$ for
the onset of the two types of modulation instability versus the trap strength
$\lambda$.  The instability lines are shown as straight lines joining
numerically determined values.
The oscillation schematics on the right (and corresponding arrows) indicate
that left-right-left oscillation modes are persistent everywhere above the
$g_{12}^a$ line, while in-out-in modes are persistent only above the higher
$g_{12}^s$ line.  The spatially symmetric instability ($g_{12}^s$ line) is the
one relevant for experimental situations with symmetric traps.  Squares mark
values used in the dynamical simulations of Figs.\ {\ref{fig:Fig1}} and
{\ref{fig:Fig2}} (Table 1).
}
\end{figure}

The two lowest-lying eigenmodes are odd modes encoding out-of-phase physics.
For $g_{12}\gtrsim 1$, their frequencies are significantly below the breathing
mode, and therefore lead to relatively slow oscillations in the spin density.
This will also be visible in the real-time dynamics presented in
Section \ref{sec_timeEvolution} (first two columns of Figs.\ \ref{fig:Fig1}
and \ref{fig:Fig2}).  The forms of the corresponding eigenvectors are shown in
the lower right of Fig.~{\ref{fig:Fig3}}.  The nature of the eigenvectors
shows that the motion related to the lowest mode corresponds to the left-right
oscillations of the two species, while the next odd mode corresponds to
spatially symmetric spin motion.  The frequencies of these two modes become
imaginary at certain values of $g_{12}$, thus leading to the onset of
modulation instabilities.  The antisymmetric mode becomes unstable at smaller
value of $g_{12}$ ($g_{12}^a\approx 1.6$ for $\lambda = 0.2$) in comparison to
the symmetric mode ($g_{12}^s\approx 2.4$ for $\lambda = 0.2$).  In a
spatially symmetric trap, there is no natural mechanism for exciting the
spatially antisymmetric mode.  On the other hand, any difference between
$g_{11}$ and $g_{22}$ naturally excites the second (spatially symmetric) mode.
Thus, the second mode, occurring at larger $g_{12}$, is the relevant
instability for understanding the dynamics observed in experiments and
explored numerically in Section
\ref{sec_timeEvolution}.

We find similar excitation spectra for trap strengths $\lambda$ spanning
several orders of magnitude.  The spatially antisymmetric mode becomes
unstable before the spatially symmetric mode, and both instabilities get
closer to 1 as the trap gets shallower.  For example, for $\lambda = 10^{-3}$
(also shown in Fig.\ {\ref{fig:Fig3}}) the lowest instability sets in for
$g_{12}^a\approx 1.02$, while the next one appears at $g_{12}^s\approx 1.05$.
The distinction between two instabilities becomes ever smaller as we go toward
a uniform system $\lambda \rightarrow 0$, where the phase-separation condition
Eq.~(\ref{eq:cond}) becomes exact.  Nevertheless, even for shallow traps, the
issue is not purely academic as the precision in experimental measurement and
control of scattering lengths continues to improve \cite{vanDruten_expt,
EgorovOpanchuk12}.

In Fig.~{\ref{fig:Fig4}} (main panel), the results of the stability analyses
are combined to present a dynamical ``phase diagram''.  The two lines show the
two instabilities ($g_{12}^a$ and $g_{12}^s$) as a function of trap strength
$\lambda$.  For very shallow traps, the two transition lines merge as
$g_{12}^s \approx g_{12}^a \approx 1$.  The lower transition line ($g_{12}^a$)
was previously introduced in Ref.~\cite{NavarroKevrekidis_PRA09}.  However,
for a trap and initial state with left-right spatial symmetry, this is not the
relevant dynamical transition line, because the first even mode only becomes
unstable at some higher $g_{12}$ value, given by the $g_{12}^s$ line.

In the next Section, we will see that spin pattern dynamics is indeed only
generated when the inter-component repulsion $g_{12}$ exceeds the second
instability line ($g_{12}>g_{12}^s$), and that crossing the first instability
($g_{12}^a<g_{12}<g_{12}^s$) is not enough for pattern formation in a
spatially symmetric trap.

\section{Dynamical features across the parameter space} 
\label{sec_timeEvolution}

\begin{figure*}[!tb]
\centering  
\includegraphics[width=0.85\textwidth]{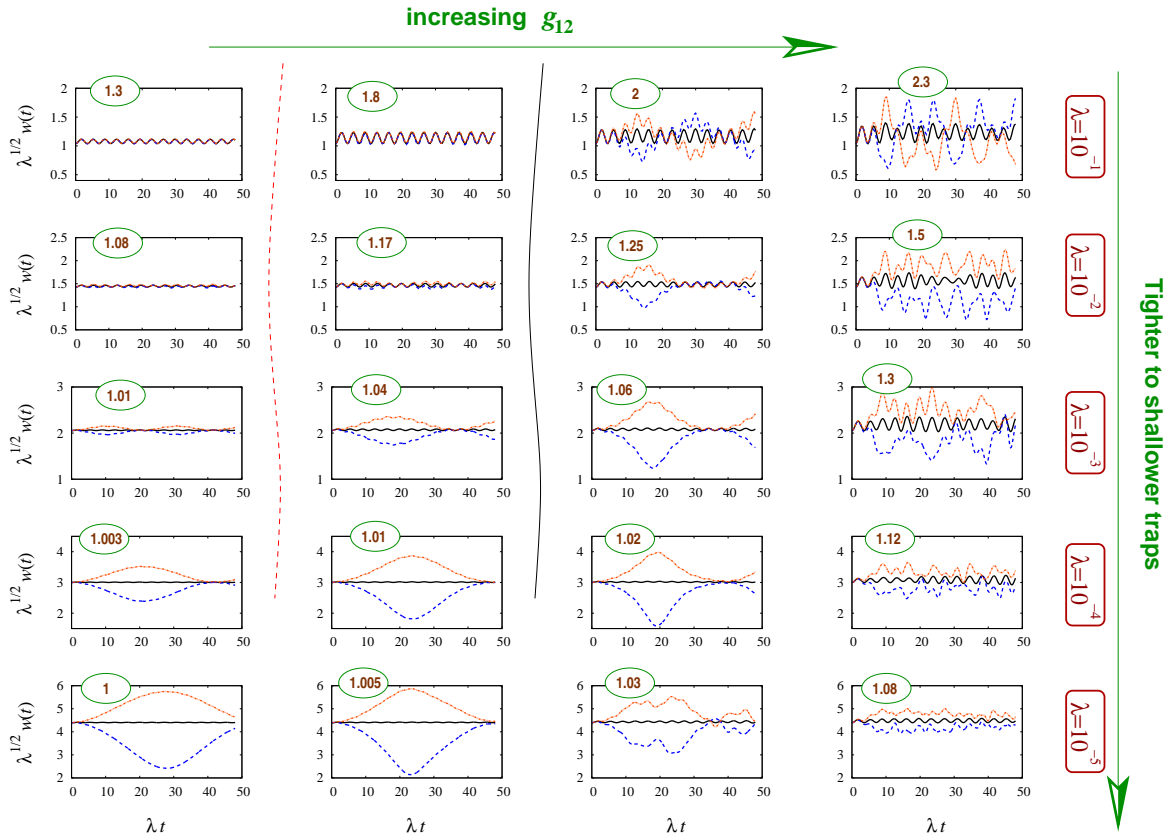}
\caption{  \label{fig:Fig1}  (Color online.) 
Time evolution of root-mean-square widths after $\pi/2$ pulse (interaction
quench).  First component width $w_1(t)$ is shown as blue dashed line, second
component width $w_2(t)$ is shown as red solid line (gray solid without
color), the total width $w(t)= \sqrt{(w_1^2(t)+w_2^2(t))/2}$ is the black
solid line intermediate between the other two.
From top to bottom: tight to shallow traps.
For each trap strength, four values of $g_{12}$ (indicated near top of each
panel) from Table \ref{tab:Tab} are used.
The two lines separating first and second column (red dashed) and second and
third column (black solid) indicate the `positions' of instability lines, from
Figure \ref{fig:Fig4}.
While the first two columns look qualitatively the same and show
regular oscillatory dynamics, in the third column we observe aperiodic motion
of stronger amplitude that we relate to the onset of spin pattern dynamics.  The
spin dynamics is even more pronounced in the fourth column.
}
\end{figure*}

In this Section we present and analyze the dynamics obtained from direct
numerical simulation of the Gross-Pitaevskii equations (\ref{eq:gp1}) and
(\ref{eq:gp2}), after the system is initially prepared in the ground state of
the situation $g_{11} =g_{22} =g_{12} =1$.  The subsequent dynamics is
performed with $g_{11} =1$, $g_{22} =1.01$, and several different values of
$g_{12}$ for each trap strength $\lambda$.

\begin{table}[!tb]
\begin{center}
\begin{tabular}{|c|c|c|c|c|c|c|}
\hline
$\lambda $ &$\frac{g_{12}^{(1)}}{\sqrt{g_{11} g_{22}}}$ &$\frac{g_{12}^{(2)}}{\sqrt{g_{11} g_{22}}}$&$\frac{g_{12}^{(3)}}{\sqrt{g_{11} g_{22}}}$ & $\frac{g_{12}^{(4)}}{\sqrt{g_{11} g_{22}}}$ &$g_{12}^{a}$ & $g_{12}^{s}$\\\hline \hline
$ 10^{-1}$ &1.3   & 1.8 & 2   & 2.3& 1.37& 1.92\\\hline
$ 10^{-2}$ &1.08  & 1.17 & 1.25& 1.5& 1.085& 1.23 \\\hline
$ 10^{-3}$ &1.01  & 1.04& 1.06& 1.3& 1.018& 1.050 \\\hline
$ 10^{-4}$ &1.003 & 1.01& 1.02& 1.12& 1.004 & 1.011\\\hline
$ 10^{-5}$ &1 & 1.005& 1.03& 1.08& $\approx$1&  $\approx$1 \\\hline
\end{tabular} 
\end{center}
\caption {Parameters from the first five columns are used for the
  plots in Figs.~\ref{fig:Fig1} and Figs.~\ref{fig:Fig2}. The
  instability values $ g_{12}^{a}$ and $g_{12}^{s}$ (introduced in
  Figs.\ \ref{fig:Fig3} and \ref{fig:Fig4} and discussed in Section
  \ref{sec_stability_analysis}) are also given for each trap
  strength.}
\label{tab:Tab}
\end{table}

\begin{figure*}[!tb]
\centering  \includegraphics[width=0.98\textwidth]{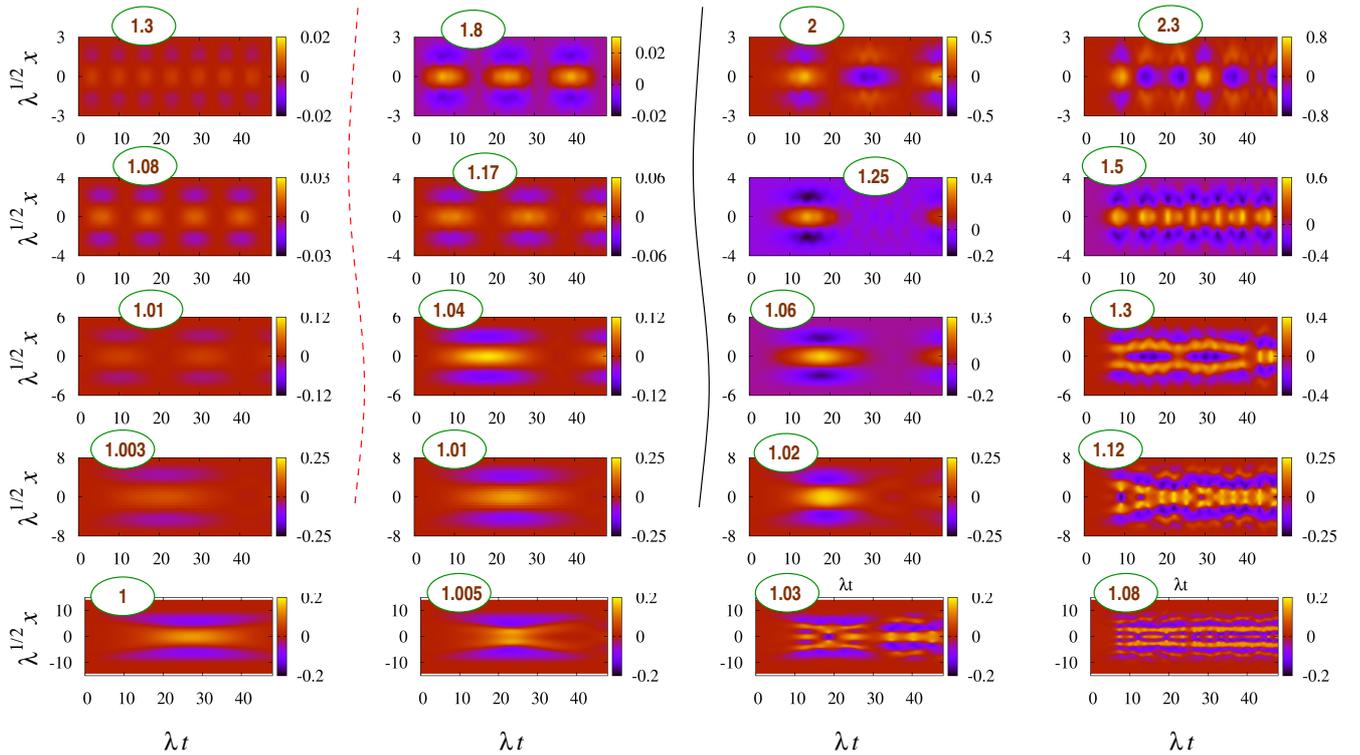}
\caption{ \label{fig:Fig2}  (Color online.) 
Spin dynamics subsequent to the  $\pi/2$ protocol, shown via the
density difference $\lambda^{-1/2}\left(|\psi_1(x,t)|^2-|\psi_2(x,t)|^2\right)$.
Traps and $g_{12}$ values are the same as in Fig.~\ref{fig:Fig1} and
Table \ref{tab:Tab}: $\lambda$ decreases from $10^{-1}$ to $10^{-5}$ from top
to bottom and $g_{12}$ values are indicated near top of each panel.
As in Fig.~\ref{fig:Fig1}, the black solid line and the red dashed line
indicate the instability lines from the ``phase diagram'' of
Fig.~\ref{fig:Fig4}.  Note the sharp change of color-scale ranges between
second and third columns in the upper rows, indicating that the dynamics
changes dramatically only across the second instability line.
}
\end{figure*}

It is difficult to show the full richness of pattern dynamics through
plots of a few quantities.  We choose to show the dynamics through two
types of plots (Figs.\ \ref{fig:Fig1} and \ref{fig:Fig2}).
Fig.~\ref{fig:Fig1} shows the  time dependence
of the root mean square widths of the two components
\begin{equation}
w_{1,2}^2(t)=\int_{-\infty}^{\infty} x^2 |\psi_{1,2}(x,t)|^2 dx , 
\end{equation}
while Fig.~\ref{fig:Fig2} shows density plots of the density
difference (spin density), $|\psi_1(x,t)|^2-|\psi_2(x,t)|^2$.  
In both figures, each row corresponds to a different trap strength
($\lambda$), and we approach the shallow trap (Thomas-Fermi) limit
going from top to bottom.  

For each $\lambda$ the four values of $g_{12}$ from Table \ref{tab:Tab} are
used for Figs.~\ref{fig:Fig1} and \ref{fig:Fig2}.  We have chosen $g_{12}$
values such that the first panel in each row is in the parameter region where
there are no instabilities, the second one is in the region where the only
instability is the antisymmetric one, and the third on each row is at $g_{12}$
values just above the second, relevant, instability.  The fourth panel on each
row is at higher $g_{12}$ values.  The choice of $g_{12}$ values with respect
to instability lines is clear in the tighter traps of the top three rows, as
also shown by squares in Fig.~\ref{fig:Fig4}.  For shallow traps (lower rows),
the instability lines are too close together and too close to $g_{12}=1$, so
making such choices is not meaningful.  
In the following, as we compare features of the different columns, we
implicitly exclude the lowest row (smallest $\lambda$).  This is also
indicated by the fact that the schematic instability lines in
Figs.~\ref{fig:Fig1} and \ref{fig:Fig2} are not extended to the lowest row.

Broadly speaking, we note that there is only regular (collective-mode) dynamics
in the second-column figures ($g_{12}^a<g_{12}<g_{12}^s$) even though an
instability is present.  There is generally a sharp difference between the
second and third figure in each row, indicating that the second instability
($g_{12}^s$) is the relevant one.  The fourth panel on each row is at higher
$g_{12}$ values, showing more rich dynamics.

In Fig.~\ref{fig:Fig1}, we show time-dependence of the individual widths
($w_1$, $w_2$) and also of the total root mean square width, $w(t)
= \sqrt{(w_1^2(t)+w_2^2(t))/2}$.  
Consistent with our observation that spatially symmetric modes (and not the
antisymmetric ones) are naturally excited in the current setup, the dynamics
shows signatures of the two most prominent spatially symmetric modes noted in
Fig.\ \ref{fig:Fig3}.  The breathing mode is the easiest to notice and most
ubiquitous --- it shows up in almost every parameter choice as
oscillations in the total density (in-phase in the two components), with a
typical period given by $2\pi/\sqrt{3}\lambda\approx 3.63/\lambda$.  This
follows from the frequency of this mode being almost constant near 
$\sqrt{3}\lambda$.

We also see out-of-phase motion of the two components, associated with the
lower spatially-symmetric mode in Fig.\ \ref{fig:Fig3}, which has odd species
parity.  In the first two columns of Fig.~\ref{fig:Fig1}, corresponding to
smaller values of $g_{12}$ such that this mode has small real frequencies,
this is excited as a regular `spin' mode.  For example, at $\lambda = 10^{-3}$
and $g_{12} = 1.04$, we observe an out-of-phase oscillation with the period of
approximately $\approx 30$, much slower than the breathing mode. In addition,
the oscillation period of the out-of-phase motion is slower in the second than
in the first column of each row, corresponding to the decreasing frequency of
the mode, as seen in stability analysis (Fig.~\ref{fig:Fig3}).  Once $g_{12}$
becomes large enough that the instability threshold for this mode is crossed,
the oscillation amplitudes increase sharply and the width dynamics becomes
strongly aperiodic and irregular (third column of Fig.~\ref{fig:Fig1}).  This
signifies the onset of pattern dynamics, as opposed to the excitation of a
regular collective mode around a stable state.  Irregularity of the width
dynamics at stronger $g_{12}$ is even more apparent in the fourth column of
Fig.~\ref{fig:Fig1}.

It is noteworthy that the spatially antisymmetric modes play no role
and do not show up in these dynamical simulations.  We see no
signature of the Kohn mode.  Nor do we see any sharp change associated
with the instability of the antisymmetric mode, i.e., there is no
sharp difference between the first two columns of Fig.~\ref{fig:Fig1}.

In Fig.~\ref{fig:Fig2} we show the dynamics of the ``spin density''
$|\psi_1(x,t)|^2-|\psi_2(x,t)|^2$.  The case of very shallow traps (last row),
resembles the data in Refs.~\cite{KasamatsuTsubota_PRL04,
KasamatsuTsubota_PRA06}.
As in Fig.~\ref{fig:Fig1}, the first two columns show regular oscillations,
corresponding to collective modes without instability.  A sharp change occurs,
not across the first instability line (between 1st and 2nd column), but
instead across the second instability line (2nd and 3rd columns), especially
for tighter traps (top three rows) where comparison with instability lines is
meaningful.  The sharp change can be noted through the color scales, which is
dramatically different between second and third columns in the upper rows.

\section{Length Scales of patterns} 
\label{sec_lengthscales}

\begin{figure}[!tb]
\centering \includegraphics[width=\columnwidth]{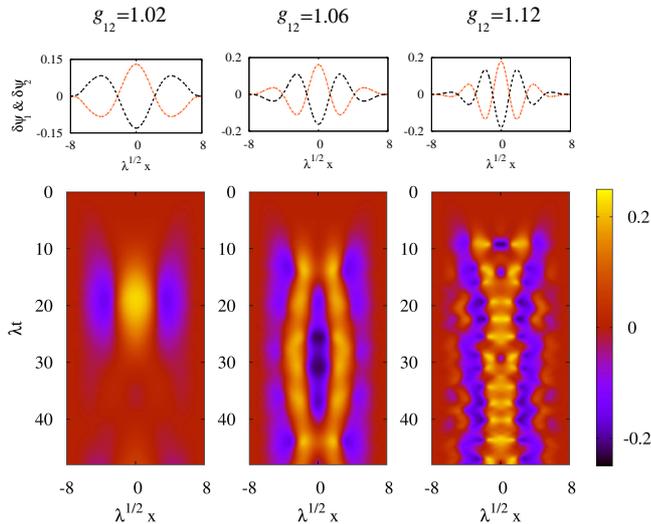}
\caption{  \label{fig:Fig5}     (Color online.) 
Top: Eigenvectors of the most unstable even eigenmodes, from the stability
 analysis of Section \ref{sec_stability_analysis}, for $\lambda = 10^{-4}$,
 and $g_{12} = 1.02$, $1.06$, and $1.12$ from left to right.  Below each
 eigenvector, the corresponding spin dynamics after the $\pi/2$ protocol
 (parameters of Section \ref{sec_timeEvolution}) is shown through the time
 evolution of $|\psi_1(x,t)|^2-|\psi_2(x,t)|^2$.
}
\end{figure}

In homogeneous stability analysis, the length scale of patterns is inferred
from the wavevector (momentum) at which an instability first occurs.  Since we
perform our stability analysis specifically for trapped systems, we do not
have a momentum quantum number.  Nevertheless, the eigenvectors of the
unstable modes contain information about the form of patterns generated in the
dynamics of the trapped system.  This is illustrated in Fig.~\ref{fig:Fig5},
where the eigenvectors of the lowest unstable even-parity modes are shown for
several values of $g_{12}$, together with the spin patterns generated in the
non-equilibrium dynamics.  There is a close match between the distance between
nodes of the eigenvectors (rough analog of `wavevector') and the length scales
involved in the patterns.

In Fig.~\ref{fig:Fig2} we see that the patterns contain more spatial structure
in shallow traps.  The top two rows (tight traps) only show in-out-in type of
patterns.  This can be understood from the idea that the interactions induce
length scales (`healing lengths') in the problem, which are smaller for larger
interactions, and which set the length scale of spatial structures.  For tight
traps, the healing length set by the interactions is large or comparable to
the cloud size, so that only global dynamical patterns are generated.  In such
traps, generation of complex patterns with many spatial oscillations would
require much higher values of $(g_{12}/g_{11}-1)$.  For shallow traps, the healing
length becomes much smaller than the cloud size; as a result one can have a
multitude of dynamical spin structures in the system, of the type seen in
experiments and prior simulations \cite{KasamatsuTsubota_PRA06,
vanDruten_expt, MertesKevrekidisHall_PRL07}.  This heuristic explanation can
be made quantitative by counting the number of nodes appearing in the
eigenmodes (as in Fig.~\ref{fig:Fig5}).

\section{Conclusions and Open Problems} \label{sec_conclusions}

In this article we have analyzed a widely used dynamical protocol for
two-component BECs, which involves starting from the ground state of one
component and switching half the atoms to a different component through a
$\pi/2$ pulse.  We have presented a stability analysis suitable to the trapped
situation, and also presented results from extensive dynamical simulations.
Through an analysis of unstable modes, we have presented a classification of
the parameter space into a number of dynamically distinct regions, in relation
to the prototypical initial state.  This may be regarded as a dynamical
``phase diagram''.

In the `stable' regime of parameter space (no modulation instabilities), our
stability analysis explains the observed slow spin oscillations compared to
the fast breathing mode oscillations of the total density.  We demonstrate
that the important ``phase transition'' line for spatially symmetric
situations relevant to most experiments is not the first instability (studied
in Ref.~\cite{NavarroKevrekidis_PRA09}), but a second transition line.  The
first instability is antisymmetric in space, and as a result is not naturally
excited in a symmetric trap.

Our stability analysis is performed relative to a stationary state of the
situation $g_{11}=g_{22}$. The $\pi/2$ pulse of the experiments (in the cases
where $g_{11}{\neq}g_{22}$) can be considered as turning on a nonzero
$(g_{11}-g_{22})$, i.e., turning on `buoyancy' such that one component gains
more energy by being in the interior of the trap compared to the other.  This
helps to select instability modes which are symmetric in space.

Since we have used a stability analysis with $g_{11}=g_{22}$ to analyze
dynamics with $g_{11}{\neq}g_{22}$, an obvious question is how the ratio
$g_{22}/g_{11}$ affects the regime of applicability of this scheme.  We expect
that features of this ($g_{11}=g_{22}$) stability analysis are useful for
dynamical predictions as long as $g_{12}/g_{11}-1$ is roughly more than
$g_{22}/g_{11}-1$.  For example, for shallow traps (small $\lambda$), the
instabilities occur at $g_{12}/g_{11}-1$ values comparable to 0.01, which is
why the placement of parameters in the three dynamical regions of the `phase
diagram' (Fig.\ \ref{fig:Fig4}) is not meaningful for the smallest $\lambda$
values (lowest rows of Figs.\ \ref{fig:Fig1} and \ref{fig:Fig2}).

For the stability analysis we used a reference stationary state which is of
course not the initial state: the initial state is the ground state for
$g_{11} =g_{22} =g_{12}$, while the reference state is the lowest-energy
spatially symmetric stationary state corresponding to the final value of
$g_{12}$.  The instability lines found in this stability analysis would
describe even better a situation where the dynamics is triggered by a small
quench of $g_{12}$, as opposed to the changes of $g_{12}$ that we consider
here, which can be relatively large.  We have looked at some examples of this
type of dynamics and indeed find instabilities matching the stability analysis
extremely well.  However, although the initial state in the $\pi/2$ dynamics
is somewhat different from the reference state of our stability analysis, our
results show that this stability analysis does provide an excellent overall
picture of the dynamics generated by the $\pi/2$ protocol.

Our work opens up a number of questions deserving further study.  First of
all, we have thoroughly explored the $\lambda$--$g_{12}$ parameter space,
while assuming that the intra-component interactions $g_{11}$ and $g_{22}$ are
unequal but close in value.  The regime of large difference ($g_{11}-g_{22}$)
clearly might have other interesting dynamical features which are yet to be
explored.  

Second, in this work we have restricted ourselves to the mean field regime.
While the mean field description captures well the richness of pattern
formation phenomena (c.f.\ Refs.~\cite{KasamatsuTsubota_PRL04,
KasamatsuTsubota_PRA06, MertesKevrekidisHall_PRL07, Ronen_PRA08,
NavarroKevrekidis_PRA09} in addition to present work), it may be worth asking
whether quantum effects beyond mean field might have interesting consequences
for the patter dynamics generated by a $\pi/2$ pulse.  For bosons in elongated
traps, regimes other than mean field (such as Lieb-Liniger or Tonks regimes)
may occur naturally in experiments \cite{PetrovShlyapnikovWalraven_PRL00,
FuchsGangardtShlyapnikov_PRL05, Weiss_tonksgas_Science04, NJvD_PRL08,
Naegerl_excited_stronglycorrelated_Science09}.  Dynamics subsequent to a
$\pi/2$ pulse in strongly interacting 1D gases outside the mean field regime
is an open area of investigation.

Third, we have assumed a spatially symmetric trap and an initial condition
with spatial symmetry, and this plays a crucial role in the selection of
instability channels.  In a real-life experiment, the trap will have some
left-right asymmetry.  Also, thermal and quantum fluctuations can initiate
spatially antisymmetric excitations.  The extent to which a small spatial
asymmetry affects spin dynamics remains unexplored; in such a case we would
have some type of competition between two types of instabilities.
Ref.~\cite{NavarroKevrekidis_PRA09} has studied dynamical effects of
fluctuations (noise), but the effects of thermal and quantum fluctuations is
yet to be studied in the context of a $\pi/2$ protocol.

Finally, one could consider time evolution and spatiotemporal patterns
generated by a $\pi/2$ pulse in the presence of an optical lattice, described
by the dynamics of a two-component Bose-Hubbard model.  This is a situation
easy to imagine realizing experimentally.  One could speculate complex
interplay between spin dynamics and the spatial arrangement of Mott and
superfluid regions.

\begin{acknowledgments}

I.V.\ acknowledges discussions with A.~Bala\v z and support by the Ministry of
Education and Science of the Republic of Serbia, under project No. ON171017.
Numerical simulations were run on the AEGrandomIS e-Infrastructure, supported
in part by FP7 projects EGI-InSPIRE, PRACE-1IP, PRACE-2IP, and HP-SEE.
NJvD acknowledges support from FOM and NWO. 

\end{acknowledgments}

\end{document}